# Global equation of state and phase transitions of the hard disc systems

**Hongqin Liu***

Integrated High Performance Computing Branch, Shared Services Canada, Montreal, QC, Canada

## Abstract

The hard disc system plays a fundamental role in the study of two-dimensional matters [1-3]. High-precision compressibility data from computer simulations have been reported for all the phases and phase transition regions [4-15]. In particular, Bernard and Krauth (Phys. Rev. Lett., **107**, 155704, 2011) [10] presented a complete and accurate picture of the phase transitions of the hard disc system with simulation results. However, thorough descriptions of the system depend on analytical equations of state (EoS) over the entire density range. While majority of EoS's published are for the stable fluid region only [1,16], few attempted the liquid-hexact transition region (Phys. Rev. Lett., **11**, 241, 1963 [17]; Phys. Rev. E. **63**, 042201, 2001 [18]; **74**, 061106, 2006 [19]). All the EoS's currently available are incapable of quantitative descriptions of the phase transitions. Here we construct a simple EoS to reproduce high-precision simulation data for all the stable liquid, liquid-hexatic transition region and hexatic phase. A global EoS is then obtained when the new EoS is smoothly united with a revisited EoS for the solid phase. Using this global equation, we are able to accurately identify all the phases and the phase transitions from the stable liquid to hexatic, then to solid phases. The liquid-hexatic transition is found to be of weak first-order, namely discontinuous in density and the Gibbs free energy while continuous in entropy and the Helmholtz free energy. The hexatic-solid transition is a continuous high-order phase transition.

* Email: hongqin.liu@canada.ca; hqliu2000@gmail.com.

## I Introduction

Two dimensional (2D) systems possess rich phase transition behaviors not seen in three dimensional systems [2,3]. The celebrated Kosterlitz-Thouless-Halperin-Nelson-Young **(**KTHNY) theory based on the 2D XY model predicts a two-stage continuous phase transition scenario via an intermediate phase known as the hexatic phase [2,3,10-12,20]. The KTHNY theory also predicts that in the liquid-hexatic transition region, the pair correlation function decays exponentially in the two phases [2,20], indicating a short-range positional order. As a result, entropy change (with respect to, w.r.t., density) is continuous.

Studies on the phase behaviors of another model system, the hard disc (HD), pioneered by Alder, Hoover and Wainwright (1962, 1963) [13,17], have also attracted great attention due to its fundamental importance [1-16]. Accurate and consistent simulation results have been reported [6-12] for equation of state (EoS) data, which makes it possible to determine phase transitions numerically. Notably, Bernard and Krauth [10] have discovered that in the HD system, the hexatic-liquid transition is of weak first-order and the solid-hexatic transition is of a continuous KT type. Consistent results have been published more recently [11-12].

The phase behaviors are currently described with discrete data points only [10-12]. For describing concise features of the transitions, analytical presentations are required so that thermodynamic properties, such as the Gibbs free energy and heat capacity, can be employed. A high-precision analytical EoS for all the phases, so-called global EoS [18], can serve the purpose. A critical challenge for an accurate EoS is in the liquid-hexatic transition region where a maximum and a minimum appear in the pressure-volume plane. All the EoS's for the stable fluid [1,16] fail beyond the liquid-hexatic transition point. The first effort of describing the region was made by Alder et al. (1963) with a so-called corrected cell model [17]. Huerta et al. (2006) [19] attempted the same and similar result has been obtained (Figure S1). Luding (2001) [18] proposed a global EoS based on a free volume model. Unfortunately, none of the above can correctly predict the maximum and minimum in the transition region (see Figure 5 below).

The goal of this work is to construct a global EoS that can accurately reproduce the compressibility data for all three phases and thermodynamic functions are hence derived. By using this new EoS, the nature of phase transitions can be identified and all phase diagrams are determined analytically.

## II Global Equation of state

For the stable liquid branch, the virial series is considered the only EoS with sound physical background, in which the virial coefficients can be calculated with high precision [1]. By assuming a linear function of the virial coefficients, $B_n = c_1 n + c_0$ for $n \geq 4$, where $c_1$ and $c_0$ are constants and $n$, the order, current author [16] was able to derive a simple and accurate Carnahan-Starling [21] type EoS:

$$Z_v = \frac{1 + \frac{1}{8}\eta^2 + \frac{1}{18}\eta^3 - \frac{4}{21}\eta^4}{(1-\eta)^2} \tag{1}$$

where $Z_v$ refers to the compressibility based on the virial series, $Z = PV/Nk_BT = P/\rho k_B T$; $P$, pressure; $V$, volume; $T$, temperature; $N$, number of particles; $\rho$, density, $k_B$, the Boltzmann constant. A detailed discussion of Eq.(1) is provided in Ref [16]. The definitions of packing fraction, $\eta$, the reduced density, $\rho^*$, reduced pressure, $P^*$, reduced temperature, $T^*$, and the reduced volume, $v^*$ are all covered in the following chain of equations:

$$Z = \frac{P\sigma^2}{\rho^* k_B T} = \frac{2P^*}{\sqrt{3}\rho^*} = \frac{\pi P^*}{2\sqrt{3}\eta} = \frac{2P^* v^*}{\sqrt{3}} = \frac{1}{T^*\rho^*} \tag{2}$$

The close packing (CP) density is $\rho^*_{cp} = 2/\sqrt{3} \approx 1.1547$ and $\eta_{cp} = \pi/2\sqrt{3} \approx 0.9069$ [1]. For developing a global EoS in Ref [18], a "free volume" term that counts for the high density (liquid-hexatic) region is introduced and a bridge function is empirically used to combine the free volume term with a EoS for the stable liquid. The final global EoS could not catch the detailed features in the transition region (Figure S1). Moreover, since the Maxwell construction was not involved the EoS is not guaranteed to produce equilibrium properties. Here we adopt the ideal of "close term" [14] for the virial EoS. Eq.(1) is a compact form of the virial serial assuming that the linear relation for $B_n$ holds for all coefficients and it has a non-physical pole at $\eta = 1$. The close term will ensure that the final EoS has a physically meaningful pole $\eta_{pole} < 1$ meanwhile the EoS applies to both the stable liquid and the phase transition region. To this end, we propose:

$$Z_{lh} = Z_v + Z_p \tag{3}$$

where the subscript "*lh*" refers to the liquid/hexatic (metastable) branch, $Z_p$, the close term with a pole, $\eta_{pole} < 1$. For the hard disc and hard sphere systems, physically meaningful pole(s) can be one or two of these physical ones: random close packing (RCP) (also known as the maximally random jammed, MRJ, packing [22-24]) for the liquid/hexatic branch, or close packing (CP) for the solid branch [1,14]. Various functional forms for the closing terms have been proposed with poles at CP, RCP ($Z_{rcp}$ or $Z_{cp}$) or other empirical values. For the HD and HS systems, these terms can be written as a generic form [1,14,25]:

$$Z_p = \frac{f(\eta)}{(1-c\eta)^{\gamma}} \tag{4}$$

where $c = 1/\eta_{pole}$, $f(\eta)$ is usually a polynomial function with some empirical parameters, and $\gamma$, a constant for which the most adopted values are greater than 1 [1,14,25], while some are equal or less than 1 [1,26]. It is easy to prove the following remark: *if an EoS is constructed involving a form of Eq.(4) for a density range up to some pole, $\eta_{pole} < 1$, the only physically correct choice is $\gamma = 1$.* We use one of the most important response functions, the heat capacity at constant pressure [25]:

$$C_p^{ex} = Z^2\left(Z + \frac{\eta\partial Z}{\partial\eta}\right)^{-1} \tag{5}$$

To simplify our proof, we apply Eq.(5) only to $Z_p$, Eq.(4), without losing generality ($Z_v$ term is eliminated as $\eta \to \eta_{pole}$ and $\eta_{pole} < 1$):

$$C_p^{ex}(\eta_{pole}) = \frac{f^2(\eta_{pole})}{[f(\eta_{pole}) + \eta_{pole}f'_{\eta}(\eta_{pole})](1-c\eta_{pole})^{\gamma} + c\gamma\eta_{pole}f(\eta_{pole})(1-c\eta_{pole})^{\gamma-1}} \tag{6}$$

Since $c\eta_{pole} = 1$, we have:

$$C_p^{ex}(\eta_{pole}) \to \infty \; as \; \gamma > 1 \tag{7a}$$

$$C_p^{ex}(\eta_{pole}) \to 0, as \; \gamma < 1 \tag{7b}$$

and

$$C_p^{ex}(\eta_{pole}) = f(\eta_{pole}), \qquad as \; \gamma = 1 \tag{8}$$

For a homogeneous system $C_p^{ex} \geq 0$ ($C_p^{ex} = 0$ for an ideal gas), as a result, the only choice that will produce physically meaningful heat capacity at the pole is $\gamma = 1$.■

For the HS and HD solid systems, $C_p^{ex}(\eta_{cp}) = D$ [25,26], therefore the EoS's for HD and HS solids proposed by Alder et al. [26], $D/(1-\rho/\rho_{cp})$, are physically sound. Similarly, for HS glass system, the EoS $Z = f(\eta)/(\eta_{rcp} - \eta)$ [25], also satisfies Eq.(8). Here for the HD liquid/hexatic branch, we assume that the pole is at RCP and Eq.(8) applies. For the function $f(\eta)$, a simple polynomial function is adopted: $f(\eta) = \sum_{i=1}^{2} b_i\eta^{m_i}$ and our final EoS for HD fluid-hexatic branch ($Z_{lh}$) reads:

$$Z_{lh} = \frac{1 + \frac{1}{8}\eta^2 + \frac{1}{18}\eta^3 - \frac{4}{21}\eta^4}{(1-\eta)^2} + \frac{b_1\eta^{m_1} + b_2\eta^{m_2}}{1-c\eta} \tag{9}$$

In Eq.(9), there are 5 unknowns: $b_1, b_2, m_1,\ m_2\ and\ c = 1/\eta_{pole}$. For determining these parameters, the following constraints are imposed: (1) Eq.(8), and we accept that the "glassy" state at $\eta_{rcp}$ has the same heat capacity as solid [27], namely: $f(\eta_{pole}) = D = 2$. (2) the final EoS, Eq.(9), can accurately reproduce the simulation results for compressibility over the entire stable liquid and hexatic transition range. (3) the EoS will satisfy the Maxwell construction such that the properties calculated will be equilibrium ones. The last constraint needs to be elaborated in detail.

The Maxwell construction (also known as the equal-area-rule) , which is equivalent to the combination of the pressure and chemical potential equilibrium conditions, was devised for vapor-liquid equilibrium (VLE) calculation with the van der Waals (vdW) EoS to determine the saturated volumes and the equilibrium pressure at given temperature. In the VLE calculations, the constants of the vdW EoS are determined by the critical pressure and temperature, then the unknowns are saturated volumes and the pressure. In the HD system, however, there is no second-order phase transition and the parameters in Eq.(9) have to be determined by applying the two constraints, (2) and (3), simultaneously. Therefore, for applying the equal-area rule, we need to determine the equilibrium pressure (there is only one here) in advance. We make use of a derivative property, rigidity, which is defined as the work required to increase the density, $\omega_T = (dP/d\rho)_T = (d\bar{G}/d\rho)_T$, where $\bar{G}$ is the Gibbs free energy. According to the fluctuation theory, rigidity is inversely proportional to an average dimensionless variance in total number of particles (*N*) at constant volume and is given by the following [28]:

$$\omega_T = \frac{N_a k_B T}{[\langle(\Delta N)^2\rangle]_{V,T}} = \frac{Z^2}{C_P^{ex}} = Z + \eta\frac{dZ}{d\eta} \tag{10}$$

where $N_a$ is the Avogadro constant. For a first-order phase transition, the densities of two phases across the coexistence region are discontinuous, and $\omega_T$ (and $C_P^{ex}$) takes non-physical (negative) values. For the stable, supercooled and superheated phases, $\omega_T$ is positive. With a continuous function, Eq.(8) requires that the $P$ and $\bar{G}$ functions exhibit a maximum (or a minimum) at $\omega_T = 0$. For a second-order phase transition, all first order thermodynamic properties are continuous, and at the critical point, density fluctuation diverges, $[\langle(\Delta N)^2\rangle]_{V,T} \to \infty]$, and $\omega_T \to 0$, $C_P^{ex} \to \infty$. For a third-order phase transition, the first order derivatives ($dZ/d\eta$, $C_P^{ex}$ et al.) are continuous, but the second order derivatives ($d^2Z/d\eta^2\ dC_P^{ex}/d\eta$, ...) diverge and so on. Rigidity also provides a tool for computing the equilibrium pressure, $P^{*e}$, from simulation data. The unstable liquid-hexatic coexistence region is defined by connecting the two points at which $\omega_T = d(Z\eta)/d\eta = 0$, and the equilibrium pressure $P^{*e}$ is invariant across the two-phase region. Therefore in the two-phase coexistence region we have:

$$\eta Z = const = \pi P^{*e}/2\sqrt{3} \qquad (11)$$

where Eq.(2) has been used. By plotting the data for $\eta Z$ [10-11] in the two-phase coexistence region (see Figure 2), we get: $P^{*e} = 7.9540$, and the equilibrium temperature $T^{*eq} = 0.10888$ $[T^* = \sqrt{3}/(2P^*)]$. The value ($P^{*e}$) is in excellent agreement with those reported by Bernard and Krauth [3,10], $P^{*e} = 9.185\sqrt{3}/2 = 7.9544$; and by Kapfer and Krauth [12], $P^{*e} = 7.950$.

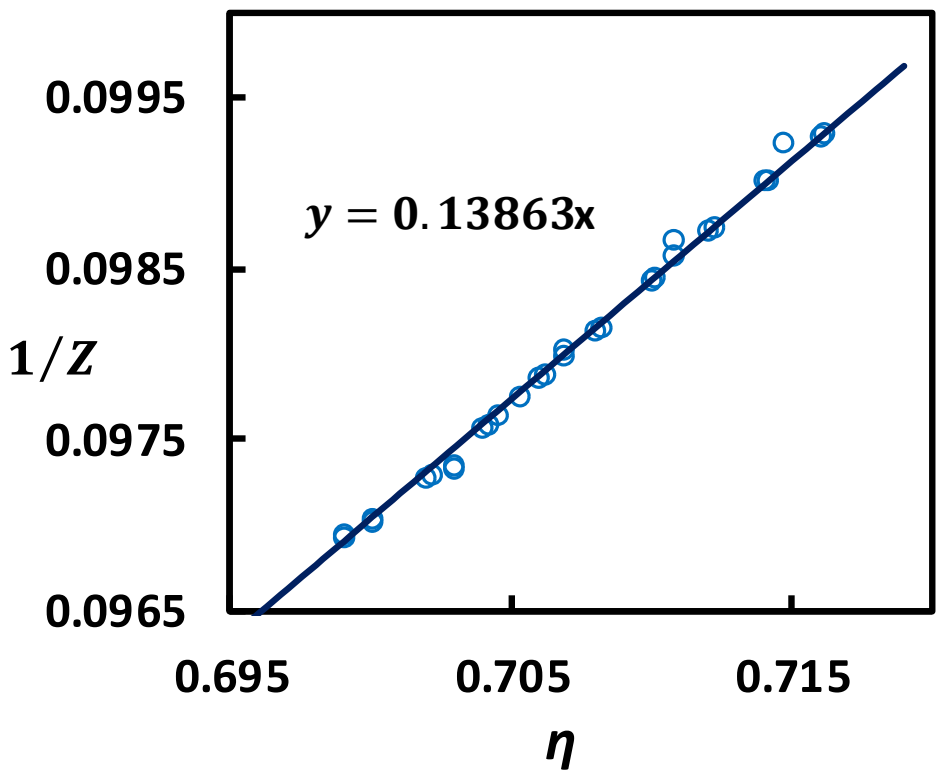

**Figure 1.** Fitting for equilibrium pressure, Eq.(11). Data sources: [10,11].

This result allows us to impose the equal-area rule as we fit the compressibility using Eq.(9). Namely the constants of the EoS will satisfy "area ABCA = area CEDC" (Figure 1a) while best reproducing the compressibility data. The equal area rule can be written as (Figure 2):

$$\int_{v_1}^{v_0} (p^{*e} - p^*)dv^* = \int_{v_0}^{v_2} (p^* - p^{*e})dv^* \qquad (12)$$

To avoid unnecessary complication of mathematical arrangements involving Eq.(9) and (12), numerical integration is used when the constraint (3) is applied. The resultant constants will provide "saturated" volumes (densities) for the liquid and hexatic phases (one pair of them in the HD system).

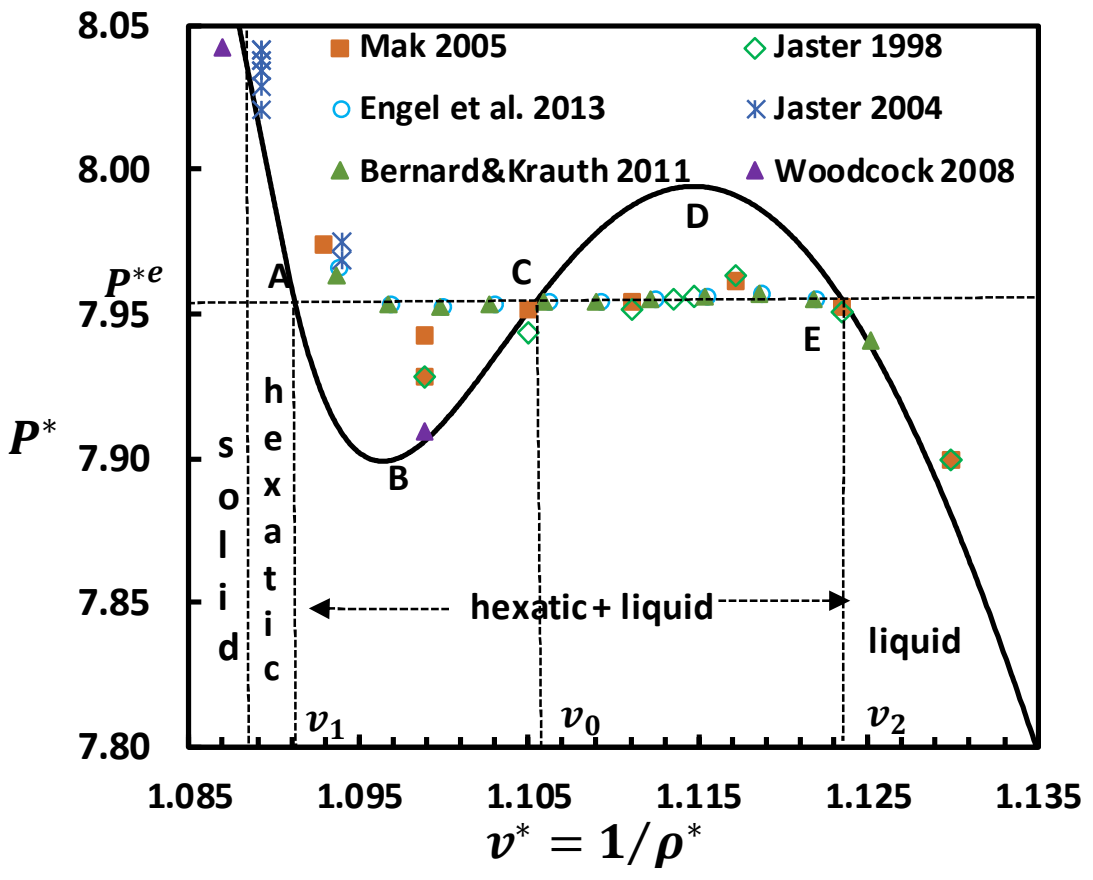

**Figure 2**. Phase diagram in the $(P, v)$ plane and the Maxwell construction. Area ABCA=area CEDC, Eq.(10). AB: super-hearted hexatic phase; DE: supercooled liquid. $P^{*e} = 7.954$. Data sources: Kolafa and Rottner 2006 [4]; Erpenbeck&Luban 1985 [5]; Woodcock 2008 [14]; Speedy&Reiss 1991 [15]; Alder et al. 1968 [26]; Mak 2006 [8]; Jaster 1999 [6]; Jaster 2004 [7]; Bernard&Krauth 2011 [10]; Engel et al. 2013 [11].

**Table 1** constants of Eq.(9)

| $b_1$ | $b_2$ | $m_1$ | $m_2$ | $1/c$ |
|---|---|---|---|---|
| $-1.04191 \times 10^8$ | $2.66813 \times 10^8$ | 53 | 56 | 0.75 |

The values obtained are listed in Table 1. The last value, $\eta_{rcp} = 1/c$, requires some further discussions. As mentioned, here a physically meaningful value should be the random close packing, $\eta_{rcp}$. This value depends on the protocol used in the simulations [22-24]. Some simulation results show that [23,24] the value of $\eta_{rcp}$ is from 0.82 to 0.84. Kansal et al. (2000) [22] employed the global bond-orientation order parameter, $\psi_6$, as a measure of the global random structure. This choice is physically sound since the hexatic state is in short-range positional order and in quasi-long range orientational order [20]. Kansal et al. [22] found that random structure ($\psi_6 \sim 0.01$) can be generated with packing fraction in the range, $0.40 \le \eta \le 0.77$. Our result from the fitting, $\eta_{rcp} = 0.75$, falls into this range. Strictly speaking, however, it should be considered as a fitting parameter.

The simple EoS, Eq.(9), works remarkably well with the constants obtained (Table 1). For the stable liquid region the accuracy for compressibility is significantly improved vs Eq.(1) with the absolute average deviation of 0.032% (21 data points [4,5], $\eta \leq 0.66$). For the liquid-hexatic transition region ($\eta = 0.66\ to\ 0.72$), high accuracy is also achieved with an AAD of 0.34% (90 data points [6-8,10-12]). Most importantly, the EoS obeys the equal-area rule in the region.

For the HD solid phase ($\eta = 0.72 \sim 0.9$) and the hexatic-solid transition region ($\eta = 0.7017 \sim 0.722$), consistent simulation data have been reported by Alder et al. [26], Speedy and Reiss [15] and more recently, Beris and Woodcock [14]. Using these data sets, we re-assessed the EoS of Alder et al. [26], and the following simple EoS is obtained:

$$Z_{solid} = 2/\alpha + 1.9 + \alpha - 5.2\alpha^2 + 114.48\alpha^4 \quad (13)$$

where $\alpha = \rho^*_{cp}/\rho^* - 1$. Eq.(13) can accurately reproduce all the compressibility data in the density range $\eta = 0.715 \sim 0.9069$ [14,15,26] with an AAD of 0.27% (30 data points). The coefficients of Eq.(13) are determined by imposing two constraints: (1) accurately reproducing the simulation data for the compressibility; (2) Eq.(13) being smoothly united with Eq.(9). Remarkably, by applying both Eq.(9) and (13) to the hexatic-solid transition region, respectively, the compressibility of the solid branch, Eq.(13), is found to be tangent to that of the hexatic branch, Eq.(7), at the "transition point" ($\eta_t = 0.720, T^*_t = 0.108717$): $Z_{lh}(\eta_t) = Z_{solid}(\eta_t) = 10.0335$; $dZ_{lh}/d\eta|_{\eta=\eta_t} = dZ_{solid}/d\eta|_{\eta=\eta_t} = 40.9$ (Figure 3). Consequently, a global EoS for all phases can be written as:

$$Z = \begin{cases} Z_{lh}, \eta < \eta_t \\ Z_{solid}, \eta \geq \eta_t \end{cases} \quad (14)$$

where $Z_{lh}$ is calculated with Eq.(9), and $Z_{solid}$ with Eq.(13).

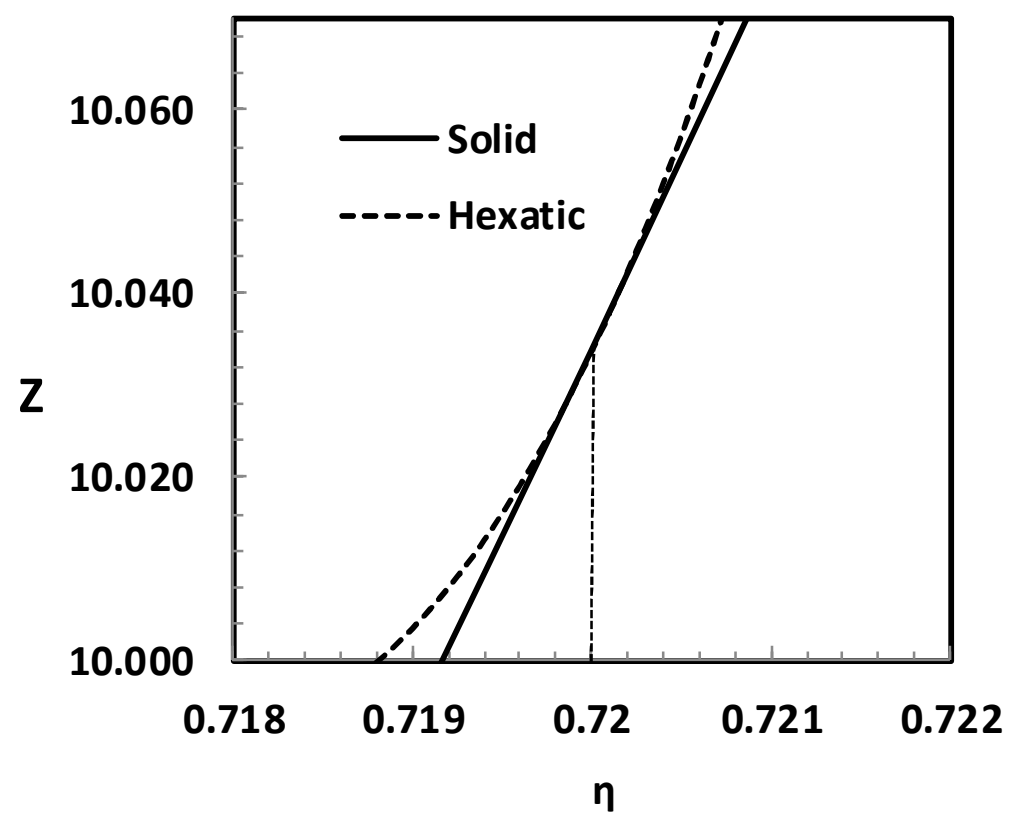

**Figure 3**. Smooth connection between compressibility of solid and hexatic phase at $\eta = 0.720$. $Z_{lh}(\eta_t) = Z_{solid}(\eta_t) = 10.0335$; $dZ_{lh}/d\eta|_{\eta=\eta_t} = dZ_{solid}/d\eta|_{\eta=\eta_t} = 40.9$.

## III Results and discussions

From Eq.(14), all other thermodynamic properties. For derivative calculations, the following equations are required:

$$\frac{dZ_{lh}}{d\eta} = \frac{2 + \frac{1}{4}\eta + \frac{1}{6}\eta^2 - \frac{103}{126}\eta^3 + \frac{8}{21}\eta^4}{(1-\eta)^3} + \frac{\sum_{i=1}^{2} b_i \eta^{m_i - 1}(m_i - ci\eta + c\eta)}{(1 - c\eta)^2} \quad (15)$$

$$\frac{dZ_{solid}}{d\eta} = \frac{dZ_{solid}}{d\alpha}\frac{d\alpha}{d\eta} = -\frac{\eta_{cp}}{\eta^2}\left(-\frac{2}{\alpha^2} + 1 - 10.4\alpha + 457.92\alpha^3\right) \quad (16)$$

To calculate various thermodynamic properties, the following integration is defined:

$$I = \int_0^\eta (Z-1)d\eta/\eta \quad (17)$$

The excess entropy, $s^{ex}$, the excess Helmholtz energy, $A^{ex}$, and the Gibbs free energy, $\bar{G}$ (chemical potential, $\mu$), can be calculated:

$$s^{ex} = \frac{S - S^{id}}{Nk_B} = -I \quad (18)$$

$$A^{*ex} = A^* - A^{*id} = I \quad (19)$$

$$\bar{G} = \mu^* = \frac{\mu}{Nk_BT} = Z - 1 + I \quad (20)$$

where $S^{id}$ is the entropy of ideal gas. From eq.(9) and (13), the integration $I$ can be carried out analytically for the liquid-hexatic branch and the solid branch, respectively:

$$I_{lh} = \frac{335\eta + 116\eta^2 + 48\eta^3}{504(1-\eta)} - \frac{673}{504}ln(1-\eta) - \sum_{i=1}^{2}\frac{b_i}{c^{m_i}}\left[\sum_{k=1}^{m_i - 1}\frac{(c\eta)^k}{k} + ln(1-c\eta)\right] \quad (21)$$

At the transition point:

$$I_{lh}(\eta_t = 0.72) = 3.8773825 \quad (22)$$

For the solid branch, we have

$$I_{solid}(\eta) \equiv I_{lh}(\eta_t) + \int_{\eta_t}^{\eta}(Z_{solid} - 1)\,dln\eta = I_{lh}(\eta_t) + \int_{\alpha}^{\alpha_t}(Z_{solid} - 1)dln(\alpha + 1) \quad (23)$$

$$Z_{solid} - 1 = 2/\alpha + 0.9 + \alpha - 5.2\alpha^2 + 114.48\alpha^4 \quad (24)$$

From Eq.(23) and Eq.(24),

$$I_{solid} = I_0 - 2ln\alpha - 110.18ln(1+\alpha) + 108.28\alpha - 54.64\alpha^2 + 38.16\alpha^3 - 28.62\alpha^4 \quad (25)$$

Finally, at the transition from Eq.(23) and (26), we obtain:

$$I_0 = I_{lh}(\eta_t = 0.72) + I_s(\alpha_t) = 3.8773825 - 2.233286 = 1.64410 \qquad (26)$$

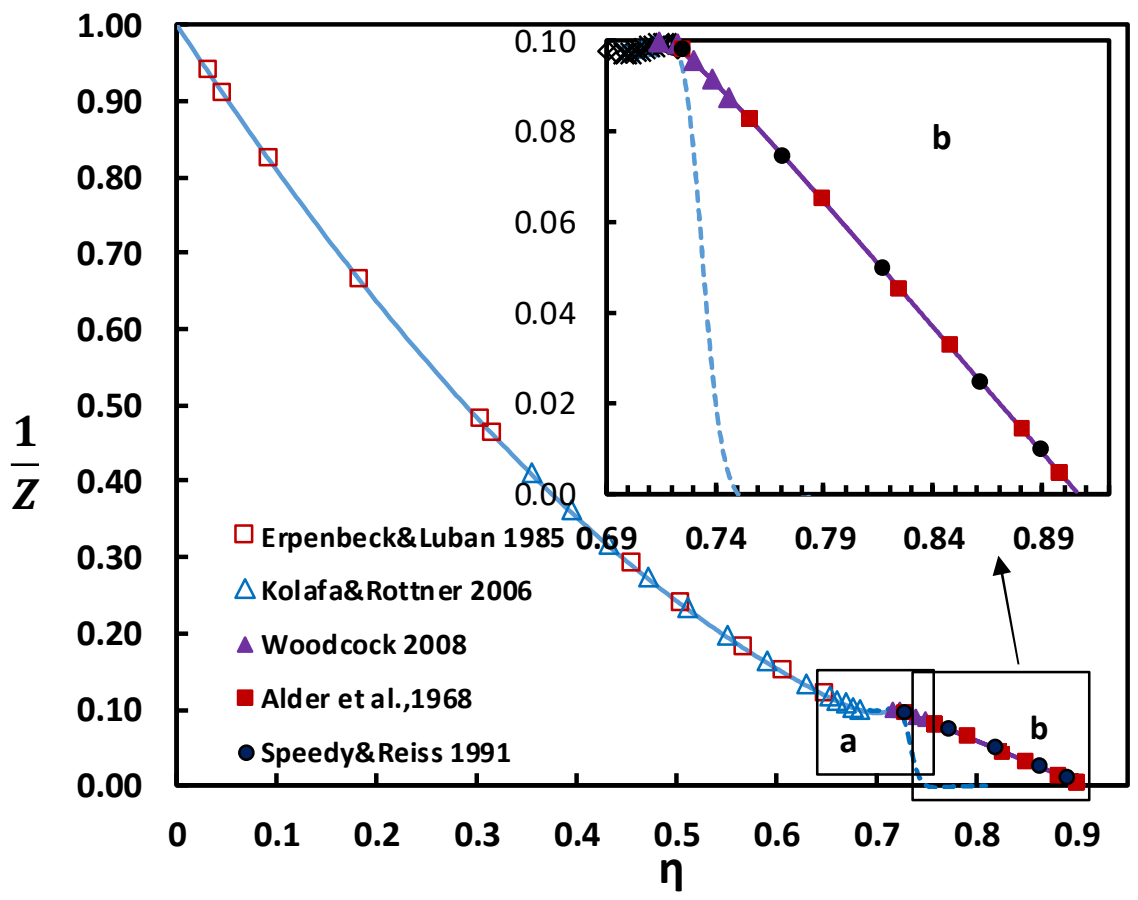

**Figure 4a**. Plots of $1/Z$ over the entire phase space. Inset b illustrates the solid phase and a pole at $\eta = 0.75$ from Eq.(9). Data sources are the same as Figure 2.

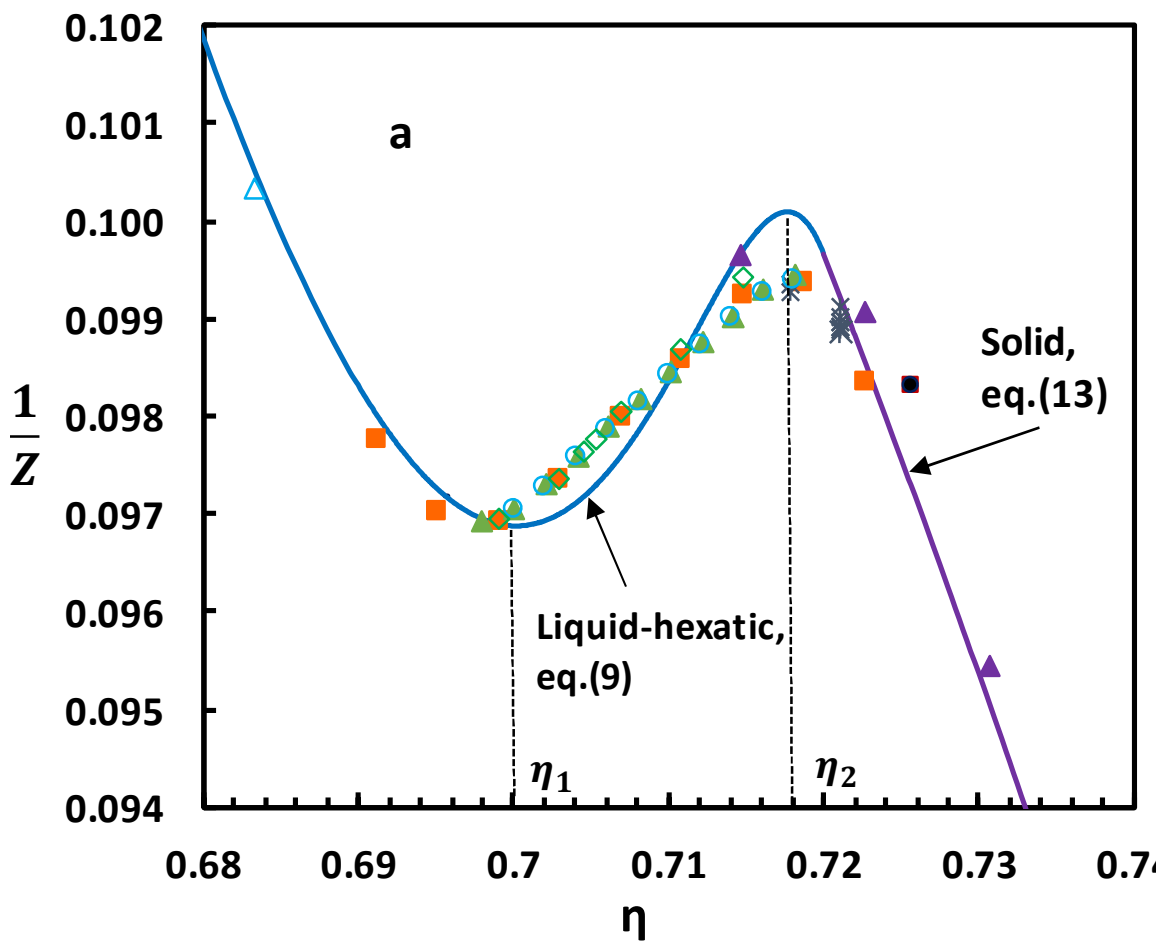

**Figure 4b**. Detailed plot of $1/Z$ in the phase transition region: enlargement of inset "a" of Figure 4a.

**Figure 4** Phase diagrams in the $(1/Z \sim \eta)$ plane.

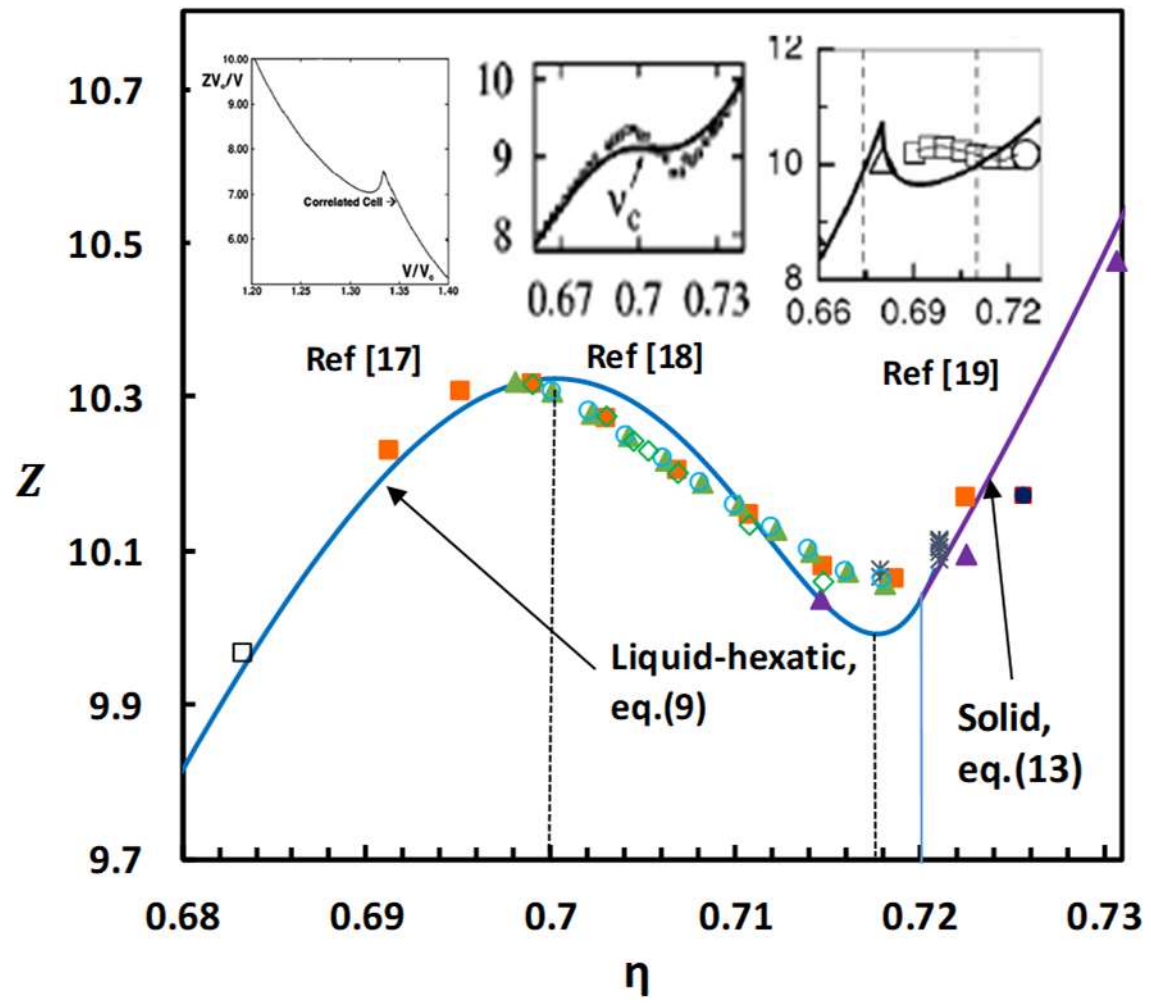

**Figure 5**. Plot of $Z \sim \eta$ in the phase transition region. The insets are adapted from Ref.[17], [18] and [19], respectively. The solid curve in the insets are EoS predictions. Noticing that the inset, Ref[17], is a plot of $ZV_0/V$ vs $V/V_0 = \eta_{cp}/\eta$ where $V_0$ is the close packing volume.

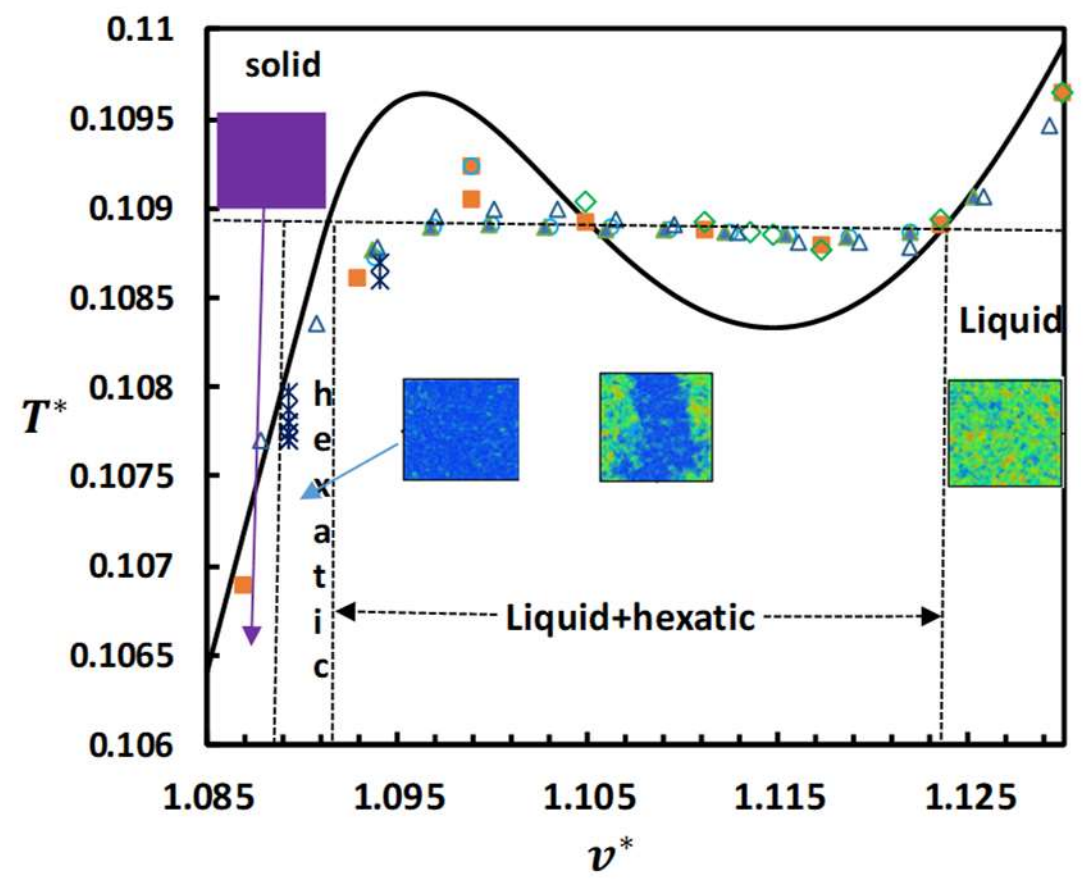

**Figure 6**. Phase diagram in the $(T^*, v^*)$ plane. The equilibrium pressure $T^{*e} = 0.108879$. The data sources are the same as listed in Figure 2. The collars of the insets of liquid and hexatic phases are adapted from Ref [3,10].

Figure 4 depicts phase diagrams in the $(1/Z, \eta)$ plane. As illustrated, the global EoS, Eq.(14), can accurately reproduce the simulation results over the entire density range from $\eta = 0$ $to$ $0.9069$. For a comparison, Figure 5 illustrates the predictions from different sources [17], [18], [19] and Eq.(14) in the $(Z, \eta)$ plane for the phase transition region. Apparently, all models proposed [17-19] are not able to quantitatively describe the phase transitions. The phase diagram in the $(T^*, v^*)$ plane is illustrated in Figure 6.

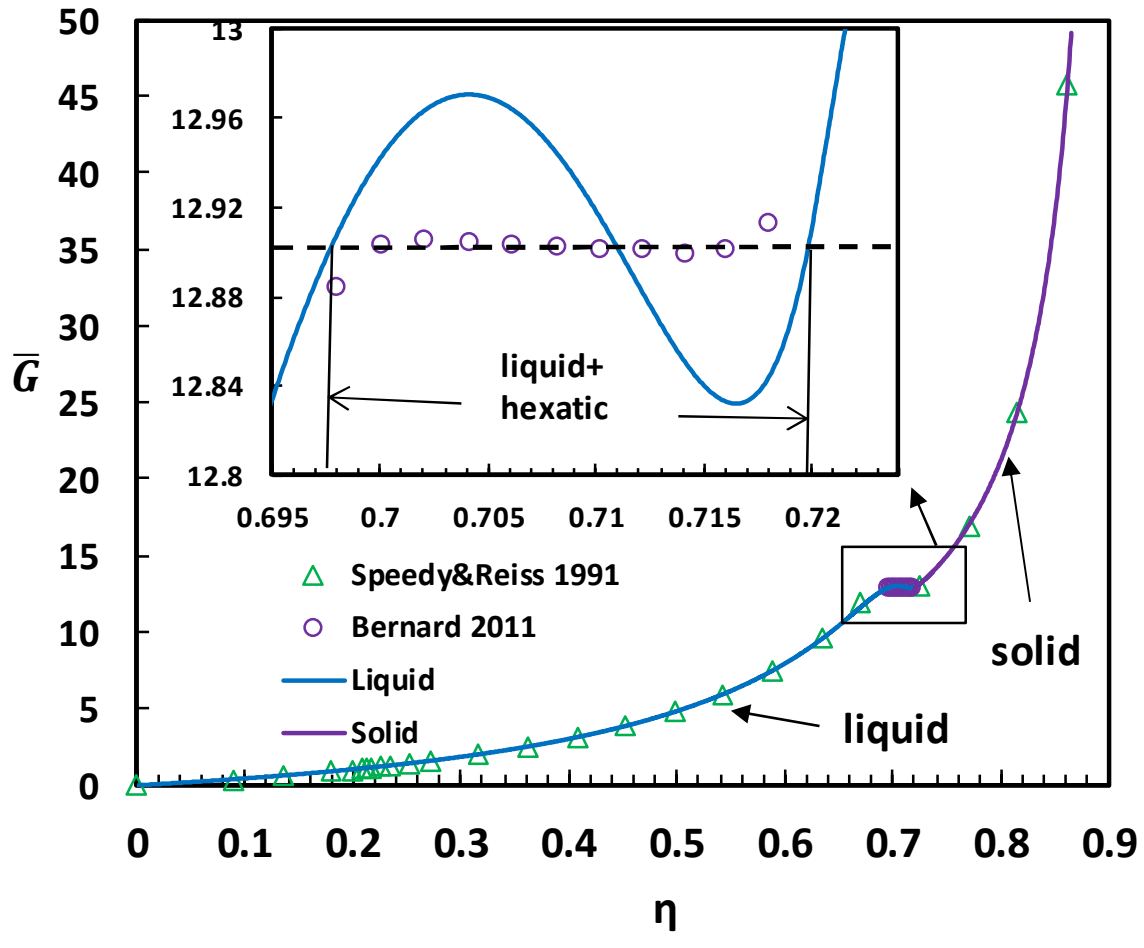

**Figure 7a.** The Gibbs free energy over the entire density range. Data sources are Speedy&Reiss 1991 [15]; Bernard 2011 [3]. $\overline{G} = \int vP(v)dv + const$, here the $const$ is adjusted from Bernard's value by $-0.015$. $\mu^{*e} = 12.901$. Bernard [3]: $\mu^{*e} = 12.916$.

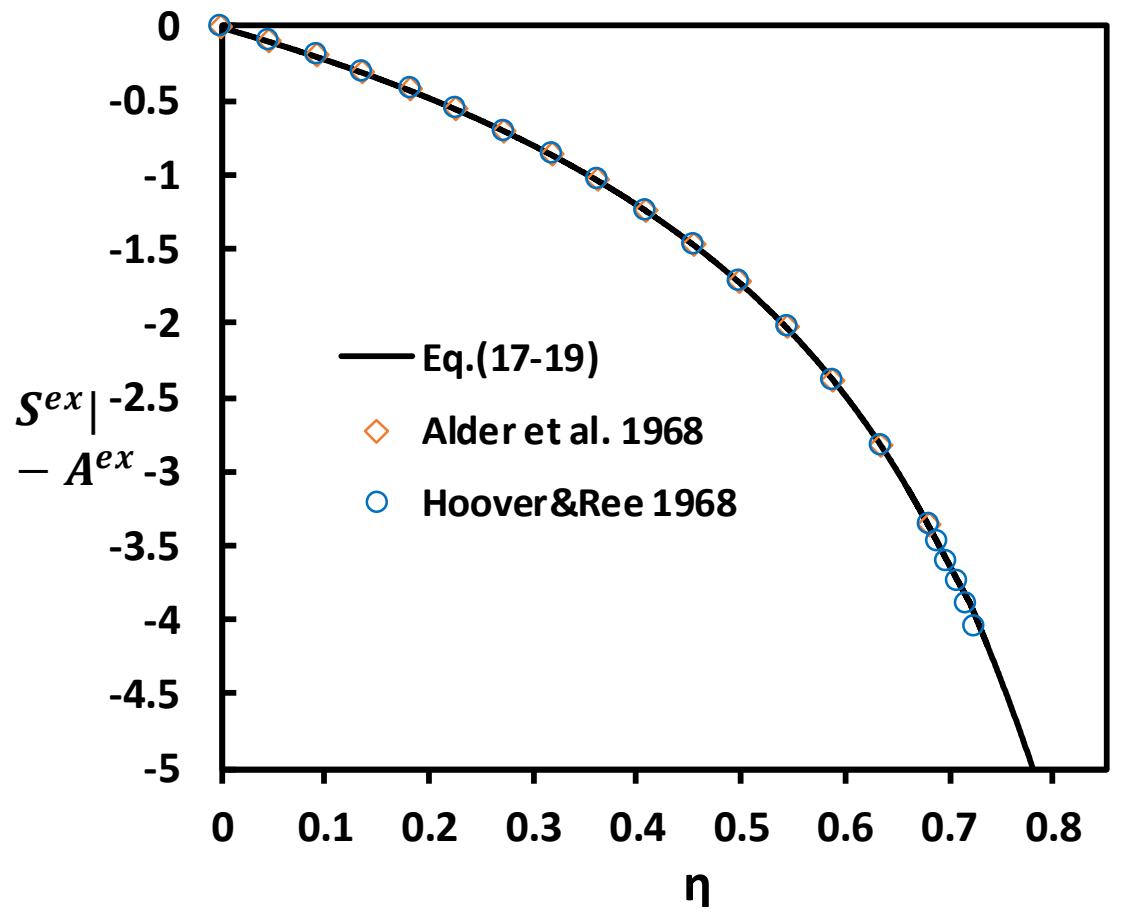

**Figure 7b**. The excess entropy and Helmholtz free energy. Data sources are Alder et al. [26]; Hoover&Ree 1968 [29].

**Figure 7**. The weak-first-order phase transition, illustrating discontinuity of the Gibbs free energy and continuity of the excess entropy and Helmholtz free energy.

Figure 7 depicts a thermodynamic consistency test by using the Gibbs free energy and excess entropy over the entire density range. The excellent agreements between predicted values and simulation data ensure that classic thermodynamics applies to all the phases and therefore, all the treatments by using the global EoS, Eq.(14), are physically sound. Figure 2a indicates that the equilibrium chemical potential is $\mu^* \approx 12.90$. Now we have a precise definition for the weak first-order phase transition, which is discontinuous in density and the Gibbs free energy while continuous in entropy and the Helmholtz free energy. The later implies that there is no latent heat.

Figure 8 illustrates the heat capacity and the hexatic-solid transition. The heterogeneous feature of the liquid-hexatic coexistence phase is reflected by the negative heat capacity. The hexatic-solid phase transition at $T_t^* = 0.108717$ $(\eta_t = 0.720)$ is a high order continuous transition.

Finally Figure 9 summarises all phases identified with the rigidity (Figure 9a), Eq.(10), and the derivative of compressibility (Figure 9b). The liquid-hexatic transition and hexatic-solid transition points are highly consistent with the simulation results [3,10,12]. In particular, the supercooled liquid and superheated hexatic regions are also identified, which can only be obtained from an analytical EoS.

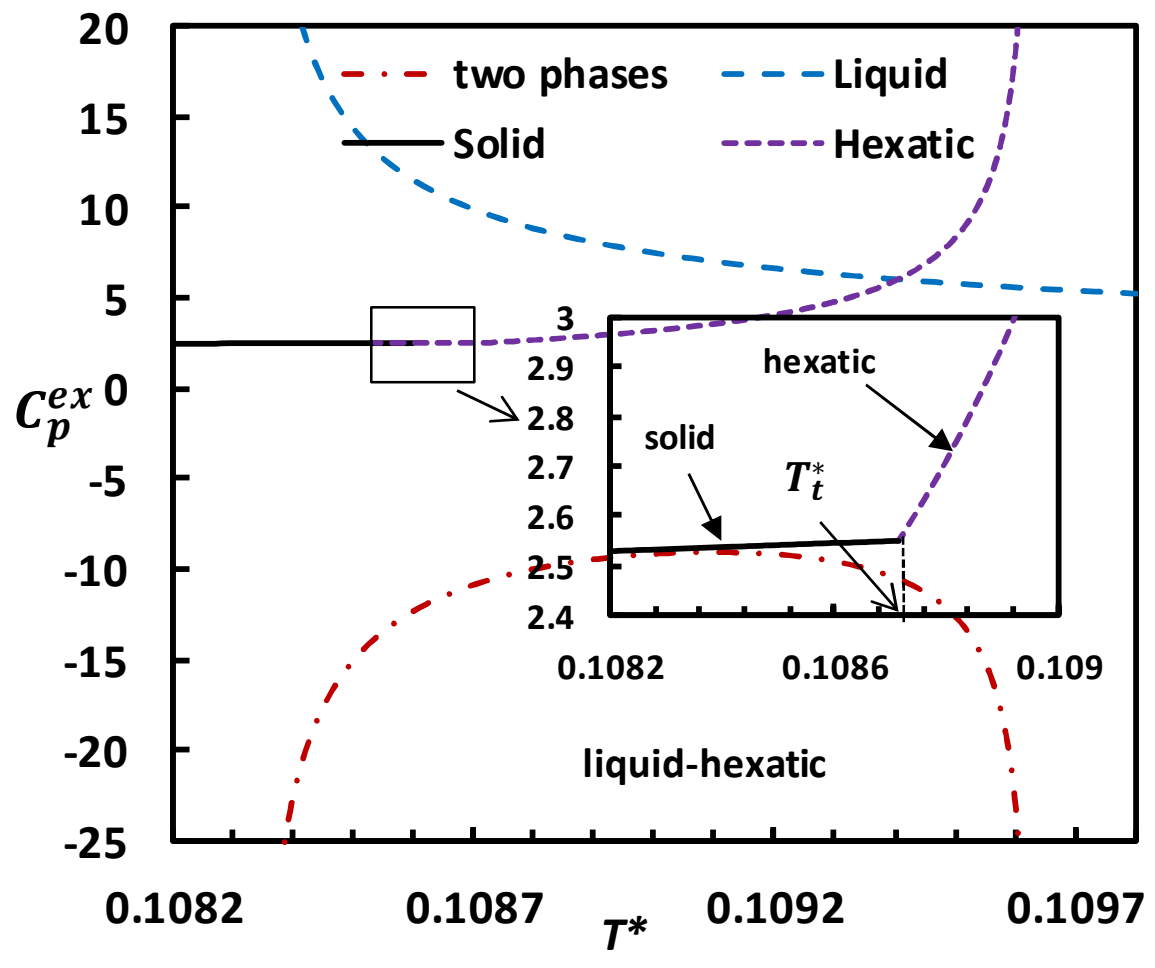

**Figure 8**. Excess heat capacity at constant pressure: $C_p^{ex} \equiv C_p/k_B - D/2$. The inset illustrates the hexatic-solid transition at $T_t^* = 0.108717$ $(\eta_t = 0.720)$. $C_p^{ex}(hexatic, T_t^*) = C_p^{ex}(solid, T_t^*)$, $dC_p^{ex}(hexatic)/dT^*|_{T_t^*} \neq dC_p^{ex}(solid)/dT^*|_{T_t^*} (< \infty)$.

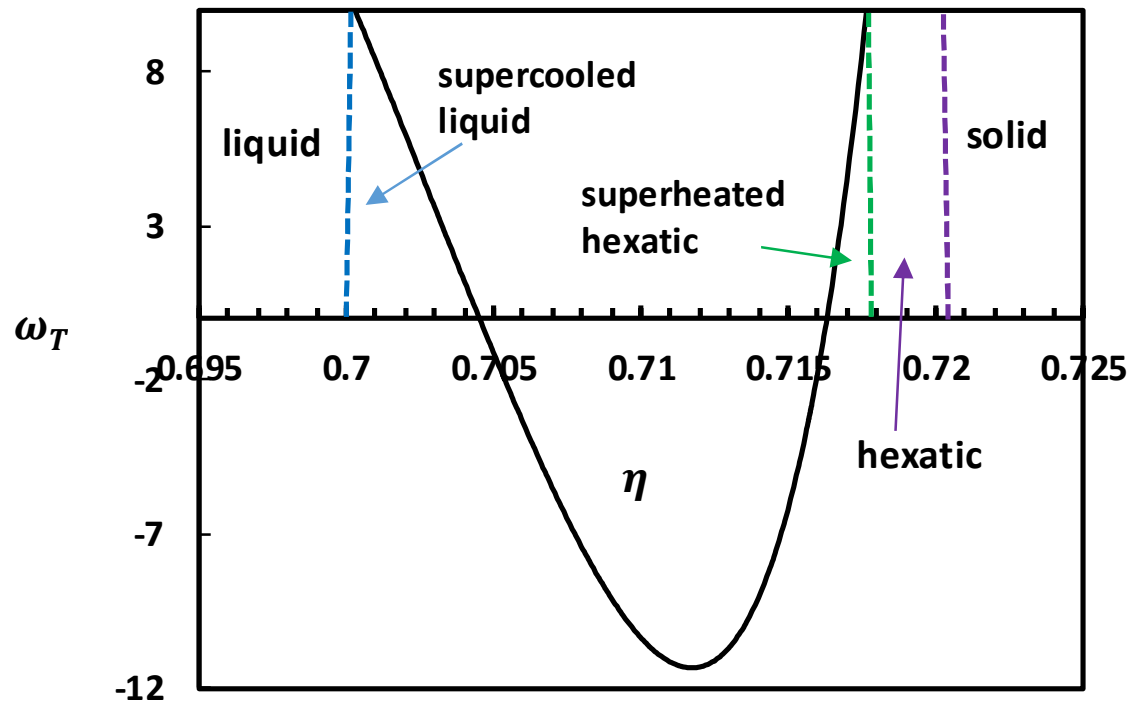

**Figure 9a**. Rigidity plot over liquid-hexatic phase transition region. Calculated by Eq.(10), Eq.(15) and Eq.(16).

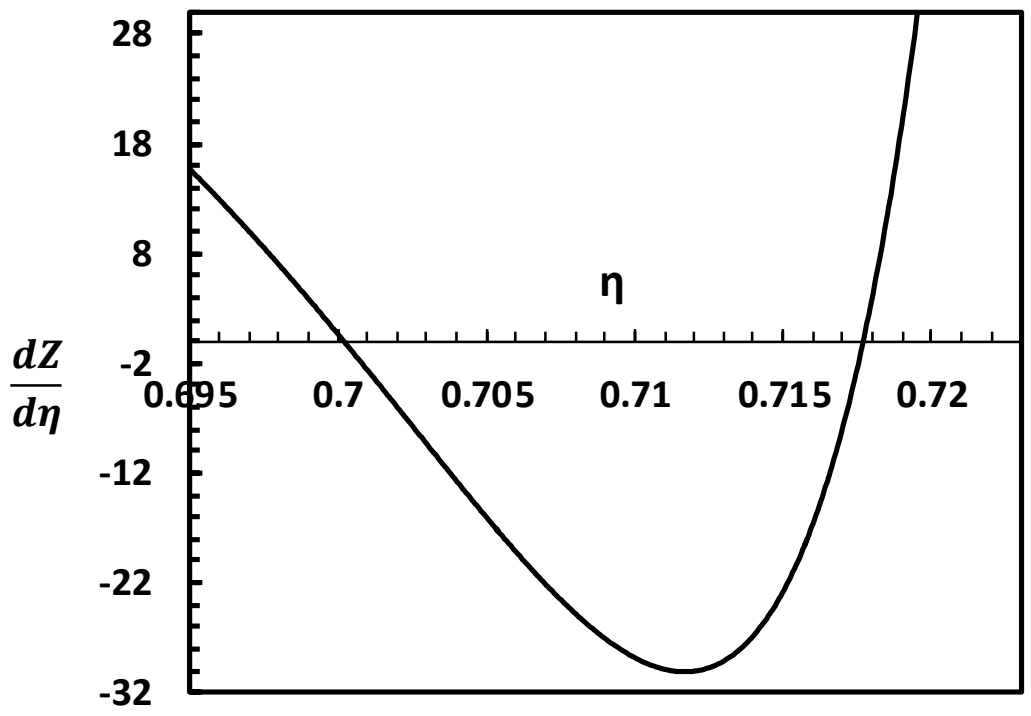

**Figure 9b**. Derivative of compressibility across the liquid-hexatic phase transition region. Calculated by Eq.(15) and Eq.(16).

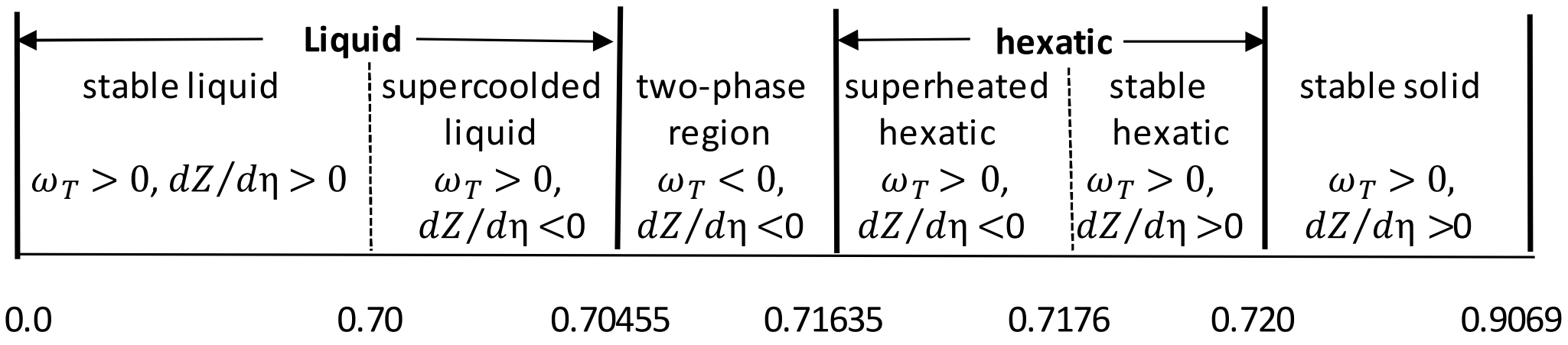

**Figure 9c.** Phases identified . Simulation results: Bernard&Krauth (2011) [3,10], liquid-hexatic: $\eta \in [0.700, 0.716]$, $\eta_{solid} \geq 0.720$; Kapfer&Krauth (2015) [12], ], liquid-hexatic: $\eta \in [0.700, 0.717]$, $\eta_{solid} = 0.7218$; hexatic [3] $\eta = 0.718$.

**Figure 9**. Summary of all phases over the entire density range.

## Conclusions

In this work, for the first time, a highly accurate and analytical EoS is proposed for the entire stable liquid phase and liquid-hexatic transition region. A global EoS is obtained when the new EoS is combined with a revisited EoS for the solid phase. By using the global EoS, both the weak first order liquid-hexatic transition and hexatic-solid transition are addressed analytically. The weak first-order liquid-hexatic transition exhibits the features of discontinuities of density and chemical potential and continuities of entropy and the Helmholtz free energy. Since at $T_t^*$, $dC_P^{ex}(hexatic)/dT^* \neq dC_P^{ex}(solid)/dT^*$ while both have finite values ($< \infty$), the hexatic-solid transition is a high order (>2) continuous transition. The phase diagrams over the entire density range from the stable liquid, supercooled liquid, two-phase coexistence region, superheated hexatic, the stable hexatic to the stable solid phases are all obtained analytically. Finally, the new EoS can also serve as a reference for developing analytical EoS for other 2D fluids, such as the Lennard-Jones fluids [1].